\documentclass[twocolumn,aps,prd,amssymb,amsmath,eqsecnum,nofootinbib]{revtex4}
\begin{document}
\title{Quantum inequalities in two dimensional curved spacetimes}

\author{\'Eanna \'E.\ Flanagan}
\affiliation{Cornell University, Newman Laboratory, Ithaca, NY
14853-5001.}

\begin{abstract}
We generalize a result of Vollick constraining the possible
behaviors of the renormalized expected stress-energy tensor of a free
massless scalar field in two dimensional spacetimes that are globally
conformal to Minkowski spacetime.   
Vollick derived a lower bound for the energy density measured by a
static observer in a static spacetime, averaged with respect to
the observers proper time by integrating against a smearing function.  
Here we extend the result to arbitrary curves in non-static
spacetimes.  The proof, like Vollick's proof, is based on conformal
transformations and the use of our earlier optimal bound in flat
Minkowski spacetime.  The existence of such a quantum inequality was
previously established by Fewster.  
\end{abstract}

\maketitle

\narrowtext

\section{INTRODUCTION AND SUMMARY}
\label{intro}

Over the last few years there has been considerable interest
in negative energies in quantum field theory.  Although energy
densities are always non-negative in classical physics, negative energies are
ubiquitous in quantum physics.  Well known examples include 
the Casimir effect \cite{Casimir1,1969PhRv..184.1272B} and squeezed
states of light \cite{1986PhRvL..57.2520W}.  In fact the energy
density measured by an 
observer at a point in spacetime can be unboundedly negative
\cite{EGJ}.  However, quantum field theory does impose non-local
constraints on negative energy effects.  
One type of such constraint are the so-called ``quantum
inequalities'', which are restrictions on the magnitude and duration of
negative energy fluxes and densities measured by inertial observers,
first introduced by Ford \cite{1991PhRvD..43.3972F,F78} and
extensively explored by Ford 
and Roman \cite{FR92,FR95,FRBH,FRWH,FRnew,Ford:1999qv} and others
\cite{Folacci:1992xg,Pfenning:1997tb,Pfenning:1998ua,Borde:2001fk,Ford:1998fa,Pfenning:1998rg,Fewster:1998pu,Fewster:1998xn,Fewster:1999kr,Fewster:1999gj,Fewster:2001js,Fewster:2002dp,Pfenning:2001wx,Verch:1999nt,Helfer:1996dj,Vollick:2000pm,Teo:2002ne,Ford2002}.

One of the main motivations for investigating these constraints is the
goal of understanding gravitational effects of negative energies as
determined by the semiclassical Einstein equation $G_{ab} = 8 \pi
\langle {\hat T}_{ab} \rangle$.  {\it A priori}, the negative energies
allowed by quantum field theory might allow the violation of cosmic
censorship \cite{Ford:1990ae,FR92}, or the production of macroscopic
traversable wormholes or closed timelike curves
\cite{Morris:1988cz,Morris:1988tu}. 
However, there has accumulated by now a considerable body of
circumstantial evidence that such objects cannot occur as solutions
to the semiclassical Einstein equation within its domain of
validity (at curvature scales below the Planck scale).
For example, Ford and Roman have used quantum inequalities to argue
that traversable wormhole solutions to the semiclassical equations
are strongly constrained \cite{FRWH}.  Planck scale
traversable wormhole solutions have been found by Hochberg et
al. \cite{Hochberg:1997ee}.

In Ref.\ \cite{Flanagan:1997gn} a particular quantum inequality was
derived for inertial observers in two dimensional Minkowski
spacetime.  This result was later generalized by Vollick to static
observers in static two dimensional spacetimes that are globally
conformal to Minkowski spacetime, using conformal transformations
\cite{Vollick:2000pm}. The purpose of this paper is to generalize
Vollick's result to arbitrary observers in non-static two dimensional
spacetimes.

We start by reviewing the results of Ref.\ \cite{Flanagan:1997gn}.
Fix a smooth, non-negative function $\rho = \rho(\xi)$ with 
\begin{equation}
\int_{-\infty}^\infty \rho(\xi) d\xi < \infty,
\label{normalization}
\end{equation}
which we will call the smearing function\footnote{In Ref.\
\cite{Flanagan:1997gn} we imposed the normalization condition 
$\int \rho d\xi =1$.  This condition is not necessary since the left
hand sides and right hand sides of the quantum inequalities
(\ref{flatT}) and (\ref{flatN}) scale linearly with $\rho$.}.
For any timelike curve 
$\gamma$ with proper time $\tau$ and normalized tangent $u^a$, we define 
\begin{equation}
{\hat {\cal E}}_{\rm T}[\gamma,\rho] = \int_{\gamma} d
\tau \, \rho(\tau) \, {\hat
T}_{ab} u^a u^b,
\label{calESdef}
\end{equation}
where ${\hat T}_{ab}$ is the stress-energy tensor.  
The quantity (\ref{calESdef}) is the 
average with respect to proper time of the energy density measured by the
observer moving along the worldline $\gamma$, weighted by the smearing
function $\rho$.  Similarly for a 
null geodesic $\gamma$ with affine parameter $\lambda$, we define
\begin{equation}
{\hat {\cal E}}_{\rm N}[\gamma,\rho] = \int_{\gamma} d
\lambda \, \rho(\lambda) \, {\hat
T}_{ab} k^a k^b,
\label{calENdef}
\end{equation}
where $k^a = (d/d\lambda)^a$.  The subscripts $T$ and $N$ denote
``timelike'' and ``null'' respectively.  Finally we define
\begin{equation}
{\cal E}_{\rm T,min}[\gamma,\rho] = \min_\omega \,\, \langle {\hat
{\cal E}}_{\rm T}[\gamma,\rho] \rangle_\omega,
\end{equation}
which is the minimum over all quantum states $\omega$ of the expected
value of the operator ${\hat {\cal E}}_{\rm T}[\gamma,\rho]$.
We define ${\cal E}_{\rm N,min}[\gamma,\rho]$ analogously.

The results of Ref.\ \cite{Flanagan:1997gn} applied to 
a free, massless scalar field $\Phi$ in two dimensional
Minkowski spacetime.  For any timelike geodesic $\gamma$ it was shown
that
\begin{equation}
{\cal E}_{\rm T,min}[\gamma,\rho] = - {1 \over 24 \pi}
\int_{-\infty}^\infty d\tau \, {\rho'(\tau)^2 \over \rho(\tau)},
\label{flatT}
\end{equation}
and for any null geodesic $\gamma$ that
\begin{equation}
{\cal E}_{\rm N,min}[\gamma,\rho] = - {1 \over 48 \pi}
\int_{-\infty}^\infty d\lambda \, {\rho'(\lambda)^2 \over \rho(\lambda)}.
\label{flatN}
\end{equation}
In this paper we consider any two dimensional spacetime $(M,g_{ab})$
that is globally conformal to Minkowski spacetime, and again consider
a free, massless scalar field $\Phi$ \footnote{We can ignore here the
well known infrared pathologies associated with massless scalar fields
in two dimensions, since our smearing functions can be taken to be of
compact support, and thus we could impose an infrared cutoff on the
theory.}. Our main results are:

(i) For an arbitrary timelike curve
$\gamma$ we have
\begin{equation}
{\cal E}_{\rm T,min}[\gamma,\rho] = - {1 \over 24 \pi}
\int_\gamma d\tau \, \left[ {{\dot \rho}^2 \over \rho}
  + \rho a^a a_a + \rho R\right],
\label{curvedT}
\end{equation}
where $a^a = u^b \nabla_b u^a$ is the acceleration of $\gamma$ and $R$
is the Ricci scalar. Here the dot denotes a derivative with respect to
proper time; we reinterpret the smearing function $\rho$ as a
function defined on the curve $\gamma$ rather than on the real line.

(ii) More generally, for any vector
field $v^a$ defined on a timelike curve $\gamma$, we have
\begin{eqnarray}
\min_\omega && \left< \int_\gamma d\tau {\hat T}_{ab} v^a v^b
\right>_\omega = -{1 \over 24 \pi} \int_\gamma  d\tau \bigg[ 
{1\over2}(\alpha^2 + \beta^2) a^a a_a 
\nonumber \\ \mbox{} && 
- (\alpha^2 - \beta^2) {\dot a}
+ {1 \over 4}(\alpha + \beta)^2 R + 2 {\dot \alpha}^2 + 2
     {\dot \beta}^2\bigg].
\label{curvedT1}
\end{eqnarray} 
Here the functions $\alpha$ and $\beta$ are defined by
\begin{eqnarray}
\alpha + \beta &=& -2 v^a u_a \nonumber \\
\mbox{} \alpha - \beta &=& 2 \epsilon^{ab} u_a v_b,
\label{alphabetadef0}
\end{eqnarray}
$\epsilon_{ab}$ is a choice of volume 2-form\footnote{The 2-form is
  normalized according to $\epsilon_{ab} 
\epsilon^{bc} = \delta_a^c$.  It can be seen that the lower bound
(\ref{curvedT1}) is independent of the choice of sign of
$\epsilon_{ab}$.}, 
\begin{equation}
a \equiv \epsilon_{ab} u^a a^b,
\label{adef}
\end{equation}
and dots denote derivatives with respect to
$\tau$.  The result (\ref{curvedT1}) reduces to (\ref{curvedT}) when
$\alpha = \beta = \sqrt{\rho}$.  

(iii) For a null geodesic $\gamma$, we have\footnote{This result was
previously noted in footnote 38 of Ref.\ \cite{Flanagan:1997er}.}
\begin{equation}
{\cal E}_{\rm N,min}[\gamma,\rho] = - {1 \over 48 \pi}
\int_\gamma d\lambda \, {\rho'(\lambda)^2 \over \rho(\lambda)},
\label{curvedN}
\end{equation}
which has the same form as the flat spacetime result (\ref{flatN}).
Here the prime denotes a derivative with respect to affine parameter
$\lambda$, where we treat $\rho$ as a function defined on $\gamma$.

We derive the results (\ref{curvedT}), (\ref{curvedT1}) and (\ref{curvedN}) in
Sec.~\ref{main} below, and discuss some implications in Sec.\
\ref{implications}.  We note that Fewster has proved that the quantity
${\cal E}_{\rm T,min}[\rho]$ is finite for any spacetime and any curve
$\gamma$ \cite{Fewster:1999gj}.  Thus, the existence of the bound
(\ref{curvedT}) follows from the very general result of Fewster.
Also, with respect to the result (\ref{curvedT1}), we note that
a quantum inequality for the time 
average of a null-null component of stress-energy was previously
derived by Fewster and Roman \cite{FR02}.

We also remark that the average along a null geodesic of any other
component of the stress-energy tensor other than ${\hat T}_{ab} k^a
k^b$ is unbounded below.  This can be seen from the flat
spacetime analysis of Ref.\ \cite{Flanagan:1997gn}.  Moreover, in two
dimensions there is no loss of generality in considering null
geodesics instead of more general null curves.  Therefore the result
(\ref{curvedN}) is the most general quantum inequality that can be derived
for null curves.

\section{DERIVATION OF THE QUANTUM INEQUALITIES}
\label{main}

The basic idea of the proof, following Vollick \cite{Vollick:2000pm}, is to
apply conformal transformations to the Minkowski spacetime results
(\ref{flatT}) and (\ref{flatN}). Our analysis differs from Vollick's
in that we allow accelerated curves.

For any metric $g_{ab}$, state $\omega$, timelike curve $\gamma$ and
weighting function $\rho$ along $\gamma$ we define
\begin{eqnarray}
I_\gamma[g_{ab},\rho,\omega] &=& 
\int_{\gamma} d
\tau \, \rho(\tau) \, \langle {\hat
T}_{ab} \rangle_\omega u^a u^b \nonumber \\
\mbox{} && + {1 \over 24 \pi}
\int_{-\infty}^\infty d\tau \, \rho \left[ {{\dot \rho}^2 \over \rho^2}
  + a^a a_a + R\right].
\label{Igammadef}
\end{eqnarray}
To derive the result (\ref{curvedT}) it suffices to show that $I_\gamma \ge
0$ always.  Consider now the 
conformally transformed metric 
\begin{equation}
{\bar g}_{ab} = e^{2 \sigma} g_{ab},
\end{equation}
where $\sigma$ is any smooth function.
Let ${\bar \omega}$ be the state on the spacetime $(M,{\bar g}_{ab})$
that is naturally associated with $\omega$ (i.e. having the same
$n$-point distributions). We define the transformed smearing function 
\begin{equation}
{\bar \rho} = e^\sigma \rho.
\label{t0}
\end{equation}

It is now a straightforward computation to show that the functional
$I_\gamma$ is a conformal invariant, 
i.e.,
that
\begin{equation}
I_\gamma[{\bar g}_{ab},{\bar \rho},{\bar \omega}] =
I_\gamma[g_{ab},\rho,\omega].  
\label{invariance}
\end{equation}
We briefly sketch the derivation.  The transformed expected
stress-energy tensor is given by 
\begin{eqnarray}
\langle {\hat T}_{ab} \rangle_{\bar \omega} &=& \langle {\hat T}_{ab}
\rangle_\omega 
+ {1 \over 12 \pi} \bigg[ \nabla_a \nabla_b \sigma -
  (\nabla_a \sigma) (\nabla_b \sigma) 
\nonumber \\ \mbox{} && 
- g_{ab} \nabla_c \nabla^c \sigma + {1 \over
    2} g_{ab} (\nabla \sigma)^2 \bigg];
\label{t1}
\end{eqnarray}
see Eq.\ (6.133) of Ref.\ \cite{BD82}.  The transformed velocity is
${\bar u}^a = e^{-\sigma} u^a$, and the transformed acceleration is
\begin{equation}
{\bar a}^a = e^{-2 \sigma} \left[ a^a + {\dot \sigma} u^a + \nabla^a
  \sigma \right].
\label{t2}
\end{equation}
Finally the transformed Ricci scalar is
\begin{equation}
{\bar R} = e^{- 2 \sigma} \left[ R - 2 \nabla_c \nabla^c \sigma\right]
\label{t3}
\end{equation}
Substituting the quantities (\ref{t0}) and (\ref{t1}) -- (\ref{t3})
together with $d 
{\bar \tau} = e^\sigma d\tau$ into the definition (\ref{Igammadef}) of
$I_\gamma$, and using 
\begin{equation}
u^a u^b \nabla_a \nabla_b \sigma = {\ddot \sigma} - a^b \nabla_b
 \sigma
\end{equation}
yields the conformal invariance
property (\ref{invariance}).

Now fix a choice of curve $\gamma$, a globally-conformally-flat metric
$g_{ab}$, a state 
$\omega$ and a smearing function $\rho$.  From the conformal invariance
property it suffices to show that $I_\gamma[{\bar g}_{ab},{\bar
\rho},{\bar \omega}] \ge 0$, where $({\bar g}_{ab}, {\bar \rho},{\bar
\omega})$ is a conformal transform of $({g}_{ab}, {\rho},{\omega})$.
Therefore, without loss of generality we can take $g_{ab}$ to be the
flat, Minkowski metric.  We can also without loss of generality take
$a^a = 0$, since any accelerated curve can be made to be geodesic
by a conformal transformation which preserves flatness
\footnote{The transformed metric will be flat if $\nabla_c \nabla^c \sigma$
vanishes, and from Eq.\ (\ref{t2}) the curve will be geodesic in the
transformed metric if $a^b a_b + a^b \nabla_b \sigma=0$ along
$\gamma$. The properties of the wave equation in flat spacetime
then guarantee that a solution satisfying this boundary condition
exists.}.  Thus, it suffices to show that $I_\gamma \ge 0$ for
geodesic curves in flat spacetime.  However, when $a^a = R =0$, the
condition $I_\gamma\ge0$ reduces to the quantum inequality
(\ref{flatT}) previously established.

An exactly analogous argument holds for null geodesics.  For any
metric $g_{ab}$, null geodesic $\gamma$ with tangent $k^a$, smearing
function $\rho$ and state $\omega$ we define 
\begin{eqnarray}
J_\gamma[g_{ab},k^a, \rho,\omega] &=& 
\int_{\gamma} d
\lambda \, \rho(\lambda) \, \langle {\hat
T}_{ab} \rangle_\omega k^a k^b \nonumber \\
\mbox{} && + {1 \over 48 \pi}
\int_{-\infty}^\infty d\lambda \, {(\rho^\prime)^2 \over \rho},
\label{Jgammadef}
\end{eqnarray}
where $k^a = (d / d\lambda)^a$.  Then one can show that
\begin{equation}
J_\gamma[{\bar g}_{ab},{\bar k}^a,{\bar \rho},{\bar \omega}] =
J_\gamma[g_{ab},k^a, \rho,\omega],
\label{invariance1}
\end{equation}
where now
\begin{eqnarray}
{\bar g}_{ab} &=& e^{2 \sigma} g_{ab} \nonumber \\
{\bar k}^a &=& e^{-2 \sigma} k^a \nonumber \\
{\bar \rho} &=& e^{2 \sigma} \rho.
\label{nullscaling}
\end{eqnarray}
Note that the conformal scaling (\ref{nullscaling}) of the smearing function
in this null case differs from the scaling (\ref{t0}) in the timelike case.
As before the conformal invariance property (\ref{invariance1}) allows
one to deduce the curved spacetime result (\ref{curvedN}) from the
flat spacetime result (\ref{flatN}).

We now turn to a proof of the result (\ref{curvedT1}).  Before
treating the case of a general vector field $v^a$ defined on the
timelike curve $\gamma$, it is useful to consider the case of a
future-pointing null vector field $k^a$ normalized according to 
\begin{equation}
k^a u_a = - {1 \over 2}.
\end{equation}
Note that there are exactly two such null vector fields along
$\gamma$, and they are expressible in terms of a choice of volume form
$\epsilon_{ab}$ by 
\begin{equation}
k^a = {1 \over 2} ( u^a - \epsilon^{ab} u_b).
\label{norm}
\end{equation}
We define
\begin{eqnarray}
&& K_\gamma[g_{ab},k^a, \rho,\omega] = 
\int_{\gamma} d
\tau \, \rho(\tau) \, \langle {\hat
T}_{ab} \rangle_\omega k^a k^b \nonumber \\
\mbox{} && + {1 \over 48 \pi}
\int_{-\infty}^\infty d\tau \, \rho \bigg[ {{\dot \rho}^2 \over
    \rho^2}
+ a^a a_a - 2 {\dot a} + {1 \over 2} R \bigg],
\label{Kgammadef}
\end{eqnarray}
where
\begin{equation}
a = \epsilon_{ab} u^a a^b = 2 k_a a^a
\end{equation}
and dots denote derivatives with respect to proper time $\tau$.
As before, it is straightforward to show that $K_\gamma$ is a
conformal invariant, in the sense that
\begin{equation}
K_\gamma[e^{2 \sigma} g_{ab},e^{-\sigma} k^a, e^\sigma \rho,{\bar
\omega}] = K_\gamma[g_{ab},k^a,\rho,\omega].
\end{equation}
Therefore, to show that $K_\gamma \ge 0$ in general, it suffices to
show that $K_\gamma \ge 0$ for geodesics in flat Minkowski spacetime.
In Minkowski spacetime, choose coordinates $(t,x)$ such that the
metric is $ds^2 = -dt^2 + dx^2$, and define $u=t+x$, $v=t-x$, and
choose $k^a = (\partial / \partial u)^a$.  Then
for the geodesic $\gamma$ given by $x=0$ we have
\begin{eqnarray}
\int_\gamma d\tau \, {\hat T}_{ab} k^a k^b \rho(\tau) &=&
\int_{-\infty}^\infty dt \, {\hat T}_{uu}(t,0) \rho(t) 
\nonumber \\ \mbox{} 
&=& \int_{-\infty}^\infty dt \, {\hat T}_{uu}(t/2,t/2) \rho(t)
\nonumber \\ \mbox{} 
&=& \int_{\tilde \gamma} d\lambda \, {\hat T}_{ab} k^a k^b 
\rho(\lambda).
\label{reduce}
\end{eqnarray}
Here the second equality follows from the fact that ${\hat T}_{uu}$
depends only on the $u$ coordinate and not on the $v$ coordinate, and
${\tilde \gamma}$ is the null geodesic $x=t$ with affine parameter
$\lambda = u$.
It follows from Eq.\ (\ref{reduce}) that the quantity $K_\gamma$
reduces to the integral (\ref{Jgammadef}) 
along the null geodesic $x=t$, which was previously shown to be
non-negative.  Thus we have proved that $K_\gamma$ is positive in
general.

Consider now a general vector field $v^a$ defined along $\gamma$.  We
fix a volume form $\epsilon_{ab}$, and define null vectors $k^a$ and
$l^a$ via
\begin{eqnarray}
\label{kdef1}
k^a &=& {1 \over 2} ( u^a - \epsilon^{ab} u_b)
 \\ \mbox{}
l^a &=& {1 \over 2} ( u^a + \epsilon^{ab} u_b).
\label{ldef1}
\end{eqnarray}
Then $k^a$ and $l^a$ are both future directed null vectors along
$\gamma$ that satisfy the normalization condition (\ref{norm}), so the
result $K_\gamma \ge 0$ applies to both $k^a$ and $l^a$.  We now
define the functions $\alpha$ and $\beta$ to be the components of the
vector $v^a$ on the basis $(k^a,l^a)$:
\begin{equation}
v^a = \alpha k^a + \beta l^a.
\label{alphabetadef}
\end{equation}
Using the definitions (\ref{kdef1}) and (\ref{ldef1}) one can invert
Eq. (\ref{alphabetadef}) to obtain Eqs. (\ref{alphabetadef0}).
We now compute the integrand on the left hand side of Eq.\
(\ref{curvedT1}):
\begin{equation}
{\hat T}_{ab} v^a v^b = \alpha^2 {\hat T}_{ab} k^a k^b + \beta^2 {\hat
  T}_{ab} l^a l^b + 2 \alpha \beta {\hat T}_{ab} k^{(a} l^{b)}.
\label{integrand}
\end{equation}
Using the identity $k^{(a} l^{b)} = - g^{ab}/4$ together with the
trace anomaly
\begin{equation}
\langle {\hat T}_a^{\ a} \rangle= {1 \over 24 \pi} R
\end{equation}
we can rewrite Eq. (\ref{integrand}) as 
\begin{equation}
{\hat T}_{ab} v^a v^b = \alpha^2 {\hat T}_{ab} k^a k^b + \beta^2 {\hat
  T}_{ab} l^a l^b - {\alpha \beta R \over 48 \pi}.
\label{integrand1}
\end{equation}
We next integrate the quantity (\ref{integrand1}) along the timelike
curve $\gamma$.  Since the coefficients $\alpha^2$ and $\beta^2$ are both
non-negative, we can apply the result (\ref{Kgammadef}) to bound the
first two terms in Eq.\ (\ref{integrand1}).  The result is the
expression (\ref{curvedT1}).  Note that the bound is optimal or sharp,
since taking the minimum over states for the first two terms in Eq.\
(\ref{integrand1}) involves the two, independent, right-moving and
left-moving sectors of the theory.

\section{IMPLICATIONS}
\label{implications}

In this section we discuss some implications of the quantum
inequalities (\ref{curvedT}), (\ref{curvedT1}) and (\ref{curvedN}) in
some specific spacetimes, and their physical interpretation.  

Consider first a uniformly accelerated or Rindler observer with
acceleration $a$ in Minkowski spacetime.  From Eq. (\ref{curvedT}) it
follows that the energy density measured by such an observer can be as
negative as $- a^2/(24 \pi)$ over arbitrarily long timescales.  This
behavior is in marked contrast to that of inertial observers, who can
only measure negative energy densities over limited timescales.
The reason for the difference can be understood by considering a state
containing a burst  
of negative energy radiation $\langle {\hat T}_{uu}(u)\rangle$
followed by a compensating burst of positive energy radiation.  An
accelerated observer can intersect the negative energy burst while
avoiding the positive energy burst if the Rindler horizon lies between
the two bursts.  A similar argument was given 
by Borde, Ford and Roman to explain the behavior of the total energy
on asymptotically null spacelike surfaces in four dimensions; see
Fig.\ 2 of Ref.\ \cite{Borde:2001fk}.

Consider next a static observer near the horizon of the static, two
dimensional black hole spacetime
\begin{equation}
ds^2 = - \left(1 - {2 m \over r}\right) dt^2 + \left(1 - {2 m \over
r}\right)^{-1} dr^2. 
\end{equation}
As shown by Vollick \cite{Vollick:2000pm}, the lower bound on the
time-averaged energy density measured by such an observer becomes
arbitrarily negative near the horizon.  From Eq.\ (\ref{curvedT}), we
can see that this effect is due to the acceleration of the static
observer, since the Ricci scalar term in Eq.\ (\ref{curvedT}) is
finite at the horizon.

Next, one can derive from the general result (\ref{curvedT1}) a
constraint on the time averaged pressure measured by an observer.  By
taking $\alpha = \sqrt{\rho}$, $\beta = - \sqrt{\rho}$ in Eq.\
(\ref{curvedT1}), one obtains 
\begin{equation}
\min_\omega  \left< \int_\gamma d\tau \rho {\hat T}_{ab} e^a e^b
\right>_\omega 
= - {1 \over 24 \pi}
\int_\gamma d\tau \, \left[ {{\dot \rho}^2 \over \rho}
  + \rho a^a a_a \right],
\label{curvedT2}
\end{equation}
where $e^a$ is any unit spacelike vector field along $\gamma$
orthogonal to $u^a$.  This is identical in form to the result
(\ref{curvedT}) except that there is no Ricci scalar term.

Finally, consider averages of energy densities over a spacelike
curve rather than over a timelike curve.  One can derive a
constraint on such averages from Eq.\ (\ref{curvedT1}) as follows
\footnote{No analogous constraints on averages of energy densities over
spacelike surfaces in four dimensions can be obtained 
\cite{Helfer:1998zs,Helfer:1999uq,Helfer:1999ur,Ford2002}.}.
Consider the transformation $g_{ab} \to - g_{ab}$.  Under this
transformation, the curve $\gamma$ becomes a spacelike curve,
$\tau$ becomes proper length rather than proper time, while the 
the set of allowed expected stress energy tensors
$\langle T_{ab} \rangle$ is invariant.  Also, the quantities $u^a$, $v^a$,
$a^a$, $\epsilon^{ab}$, $a = \epsilon_{ab} u^a a^b$, and ${\dot a}$
are even under the transformation, while the quantities $a_a$, $u_a$,
and $R$ flip sign.  We therefore obtain that
\begin{eqnarray}
\min_\omega && \left< \int_\gamma ds \, {\hat T}_{ab} v^a v^b
\right>_\omega = -{1 \over 24 \pi} \int_\gamma  ds \bigg[ 
-{1\over2}(\alpha^2 + \beta^2) a^a a_a 
\nonumber \\ \mbox{} && 
- (\alpha^2 - \beta^2) {\dot a}
- {1 \over 4}(\alpha + \beta)^2 R + 2 {\dot \alpha}^2 + 2
     {\dot \beta}^2\bigg],
\label{curvedS1}
\end{eqnarray} 
where $\gamma$ is a spacelike curve with proper length parameter $s$,
tangent $t^a = (d / d s)^a$, and acceleration $a^a = t^b \nabla_b
t^a$.  Also dots denote derivatives with respect to $s$, and $\alpha$,
$\beta$ and $a$ are defined by [cf.\ Eqs.\ (\ref{alphabetadef0}) above]
\begin{eqnarray}
\alpha + \beta &=& 2 v^a t_a, \nonumber \\
\mbox{} \alpha - \beta &=& - 2 \epsilon^{ab} t_a v_b, \nonumber \\
\mbox{} a &=& \epsilon_{ab} t^a a^b.
\label{alphabetadef01}
\end{eqnarray}
By taking $\alpha = \sqrt{\rho}$, $\beta = - \sqrt{\rho}$ we obtain 
\begin{equation}
\min_\omega  \left< \int_\gamma ds \rho \,{\hat T}_{ab} n^a n^b
\right>_\omega 
= - {1 \over 24 \pi}
\int_\gamma ds \, \left[ {{\dot \rho}^2 \over \rho}
  - \rho a^a a_a \right],
\label{curvedS2}
\end{equation}
where $n^a$ is the unit, future directed normal to the spacelike curve
$\gamma$.  As an example, consider the static spacetime 
\begin{equation}
ds^2 = e^{2 \sigma(x)} ( -dt^2 + dx^2).
\end{equation}
Then the Killing energy of a state on the hypersurface $t=0$ is
\begin{equation}
\int ds \, \langle {\hat T}_{ab} \rangle \, n^a \xi^b,
\end{equation}
where $ds = e^\sigma dx$, $\xi^a = (\partial / \partial t)^a$ is the
Killing vector field and $n^a = e^{-\sigma} \xi^a$ is the unit normal.  
From Eq.\ (\ref{curvedS2}) with $\rho = e^\sigma$ it follows that the
lower bound on this 
Killing energy is  
\begin{equation}
-{1 \over 24 \pi} \int dx \left( {d \sigma \over d x} \right)^2.
\end{equation}

\section{CONCLUSION}

We have derived very general constraints on averages of
components of 
the stress energy tensor along timelike, null and spacelike curves in two
dimensional spacetimes.  The bounds are all optimal and are expressed 
in terms of covariant quantities.  Unfortunately the methods used here
do not generalize to the more interesting case of four dimensions.

\acknowledgments
I thank Chris Fewster for helpful
discussions, and the Erwin Schr\"odinger Institute for Mathematical
Physics in Vienna, where this paper was completed, for its
hospitality.  I thank Robert Wald for suggesting to use a conformal
isometry of flat 
spacetime to map an accelerated worldline onto a geodesic worldline.
I thank Eleni-Alexandra Kontou for pointing out an error in an earlier version
of this paper.
This research was supported in part by NSF grants PHY-9722189 and
PHY-0140209 and by the Sloan Foundation.


\end{document}